\documentclass[sigconf]{acmart}
\usepackage{subfig}

\AtBeginDocument{%
  \providecommand\BibTeX{{%
    \normalfont B\kern-0.5em{\scshape i\kern-0.25em b}\kern-0.8em\TeX}}}

\setcopyright{acmcopyright}
\copyrightyear{2021}
\acmYear{2021}
\acmDOI{10.1145/3464327.3464349}

\acmConference[Mindtrek '21]{Academic Mindtrek
2021}{June 1--3, 2021}{Tampere/Virtual, Finland}
\acmBooktitle{Academic Mindtrek 2021 (Mindtrek '21), June 1--3, 2021,
Tampere/Virtual, Finland}
\acmPrice{15.00}
\acmISBN{978-1-4503-8514-5/21/06}

\begin{document}

\title{Quantum Game Jam -- Making Games with Quantum Physicists}

\author{Annakaisa Kultima}
\email{annakaisa.kultima@aalto.fi}
\affiliation{%
  \institution{Aalto University}
    \country{Finland}
}

\author{Laura Piispanen}
\email{laura.piispanen@aalto.fi}
\affiliation{%
  \institution{Aalto University}
      \country{Finland}
}

\author{Miikka Junnila}
\email{miikka.junnila@aalto.fi}
\affiliation{%
  \institution{Aalto University}
      \country{Finland}
}

\renewcommand{\shortauthors}{Kultima, Piispanen, Junnila}

\begin{abstract}
In this paper, we explore Quantum Game Jam (QGJ) as a method for facilitating interdisciplinary collaboration and creating science game prototypes. QGJ was a series of game development events, science game jams, organized five times 2014-2019. In these events game makers and quantum physicists created games about quantum mechanics, games for quantum research and games utilizing quantum computers. QGJ has worked as a platform for multidisciplinary and collaborative exploration and learning: through creating 68 game prototypes throughout the years the participants have networked and learned from each other. In addition to this, at least three prototypes have been taken into further development. In this paper we evaluate how the format of QGJ performed. We discuss both the organizing of the events as well as the utility of the game prototypes. In addition to our observations on the jams, we have evaluated the game submissions of all events (2014-2019) and gathered survey data from the participants of the fifth event (2019).  
\end{abstract}




\maketitle

\section{Introduction}
For the past decade, game jams \cite{kultima15} have grown into fruitful events for people with different backgrounds and interests to meet and create together \cite{fowler13, kultima18, kultima16b, laiti}. The current landscape of game jams is vast. There are constantly open game jams run by various actors - companies, organizations, educational institutes and private persons. For instance, a popular game jam platform and an indie game marketplace Itch.io hosts over 90 game jams at the time of the writing of this paper \cite{itch21}. 

There are different kinds of game jams varying in terms of the duration, design constraints, participant numbers and location \cite{kultima21}. Many game jams are modelled after the Global Game Jam (GGJ) or Ludum Dare (LD), the two longest running large-scale game jams. Also other game jams have had an impact on the development of the field, such as Amnesia Fortnight (commercial game jams), Train Jam (unexpected locations) or 0h Game Jam (extreme duration), to name only a few. Altogether these game creation events can take different forms and facilitate collaboration between different people, affording spaces for co-creation, networking and creativity.

Game jams as a format has been studied from various viewpoints. For instance how the creative process is facilitated \cite{zook13, kultima11, ho14}, how game jams are part of prototyping cultures \cite{musil19,yamane13}, how they can teach about accessibility issues \cite{scott13}, how jams can be used for learning \cite{preston}, and what are the social dynamics in these events \cite{turner13, reng13}. Game jams have been defined as “accelerated opportunistic game creation events where a game is created in a relatively short time frame exploring given design constraint(s) and end results are shared publicly” \cite{kultima15}. In this paper, we are exploring a special subcategory of game jams that we call \textit{science game jams}. A science game jam is a game jam where the collaboration between game makers and scientists is facilitated in order to produce games contributing to scientific work, directly (such as helping to solve research questions) or indirectly (such as building awareness or teaching a scientific topic). 

Games as well as other forms of popular media have been inspired by the world of quantum physics for a long time. For instance at Valve's game marketplace Steam, the search term 'quantum' resulted in 227 games within the over 50k game titles library at the time of the writing of this paper \cite{steam21}.  In many games, quantum physics is thinly attached to the game and is not contributing to scientific work directly or indirectly. However, it is not rare to consult experts, such as physicists in the production of commercial games. For example, in Remedy Entertainment’s \textit{Quantum Break}, the game has been built around the inspiration from actual physics \cite{kamen}.

Quantum games, where the rules and actions rely on quantum physics theory and representing actual quantum physics phenomena have mostly been presented as small projects created by physicists \cite{cantwell19,goff06,gordon10,gordon12,wootton17}. Some productions have involved professional game developers creating citizen science games \cite{heck18,sorensen,sorensen2,bell}. Citizen science games are games aimed at the public to create data through gameplay that researchers can use in solving their research questions \cite{foldit, eyewire, cooper, smanis}. In our conception, citizen science games are a subcategory of science games.

One of the recent developments impacting quantum games is quantum computing. Quantum computing has its roots in the 1980’s \cite{benioff, deutsch, feynman, manin} and the first computers accessible to general audiences were provided by the IBM Q Experience in 2016 \cite{IBMQ}. The first known game for a quantum computer was a variant of \textit{Rock-Paper-Scissors}, \textit{Cat-Box-Scissors} (2017) \cite{woottonhistory}. After this, a handful of games using a quantum computer have been developed to take advantage of quantum mechanical phenomena such as superposition and entanglement \cite{woottonhistory, becker}. Games have not only exhibited the functionalities of these technologies, but also contributed to the development and research of quantum technologies \cite{wootton18,wootton20}.

While quantum physics can be inspiring to game creators and an interesting challenge for research, due to its unintuitive nature, it can be challenging to grasp sometimes even for quantum physicists themselves. This makes quantum physics hard to communicate with traditional outreach measures and challenging to engage outsiders' attention. Also game making requires specialized expertise that takes years to develop. It is rare that someone masters both disciplines. There is a need to facilitate collaboration and knowledge change between the actors of these fields.

Game jams that aim for facilitating the collaboration of various disciplines, like science game hackathons \cite{biotech,aigames, cern}, serious game jams and hackathons \cite{megumi,alhadeff,techfugees, segap} and academic game jams or game jams for research \cite{deen14,ramzan16, cook15, goddard14} have been organized \cite{fowler13l}. The potential of game jams as part of scientific work has been deemed promising in interdisciplinary collaboration \cite{cook15, ramzan16}. However, detailed academic studies on how game jams can successfully facilitate collaborative design processes between game makers and scientists seem to be hard to find.  

Our exploration of quantum games as science games has been facilitated in the Quantum Game Jam (QGJ) project. QGJ was an intensive, over-the-weekend game jam where aspects of quantum mechanics were translated into games. In this paper, we will explore the lessons learned from these events and deepen our understanding on how the method of game jams can facilitate multidisciplinary work and work as a platform for creating new science games.

\section{STUDYING THE METHOD OF QUANTUM GAME JAMS}
\subsection{Research Questions}
In this study, we were interested in the method of game jams from two perspectives: 1) How the game jam events work as a format for interdisciplinary collaboration between game makers and scientists, and 2) How the game jam events work as a method for producing science game prototypes.

In order to explore these questions, we focused on a specific science game jam, Quantum Game Jam (QGJ) organized five times between 2014 and 2019. We devised a post-event survey of one of the events (2019), and analysed the games produced in all of the events (2014-2019). In addition to these, one author of this paper observed all five events, and one observed the fifth event.

The QGJs were not specifically set as a research experiment and the events evolved throughout the years, so we will first explain the setting, i.e. how the events were organized, what were the goals of the events, how the events varied in terms of their organizing and who participated in these events. Secondly we will explain how the survey was conducted, and thirdly how the game analysis was conducted. In the results section, we weave the findings together using our observations of the events (2014-2019) to further reflect on the research questions.

\subsection{Setting: Quantum Game Jams}
The initiative of QGJ was built as a collaboration between the Finnish Game Jam organization, University of Turku, Aalto University and University of Tampere. The initial agenda of the project was to bring quantum physicists and game experts together and see if this collaboration would give birth to successful science games - either in the form of prototypes made in the events, or as part of a longer term collaboration facilitated by learning that the events would afford. The events had various focuses depending on the given years, but in general the goal was to foster the creation of design ideas and prototypes for quantum physics citizen science projects, educational games on quantum, and later also games using quantum computers.

\subsubsection{A Series of Game Jams: Five Quantum Game Jams}
\begin{figure}
\subfloat[]{\includegraphics[width = 1.5in]{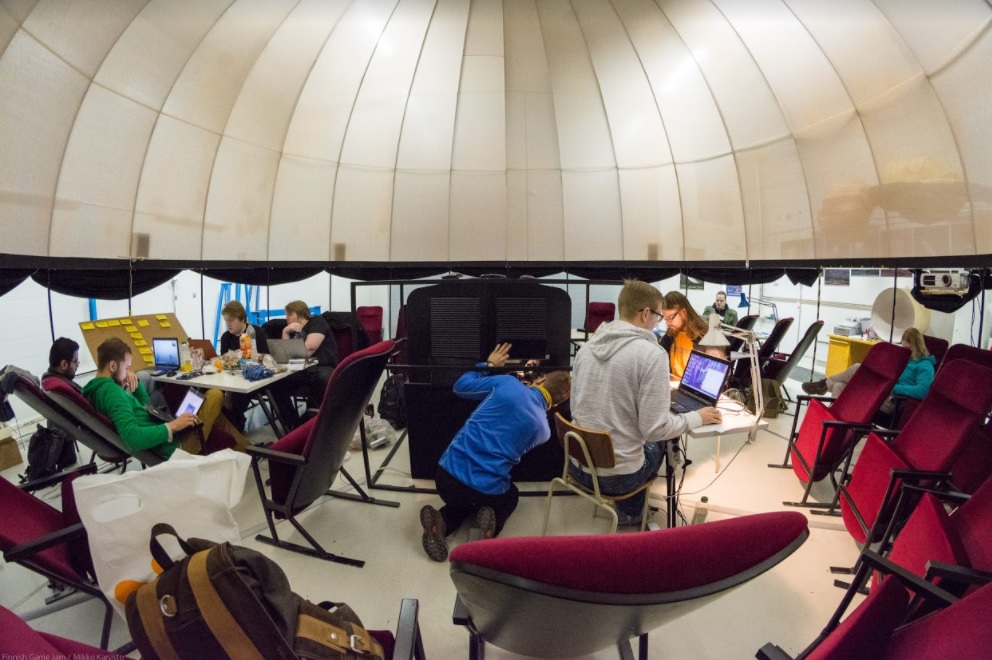}} 
\subfloat[]{\includegraphics[width = 1.5in]{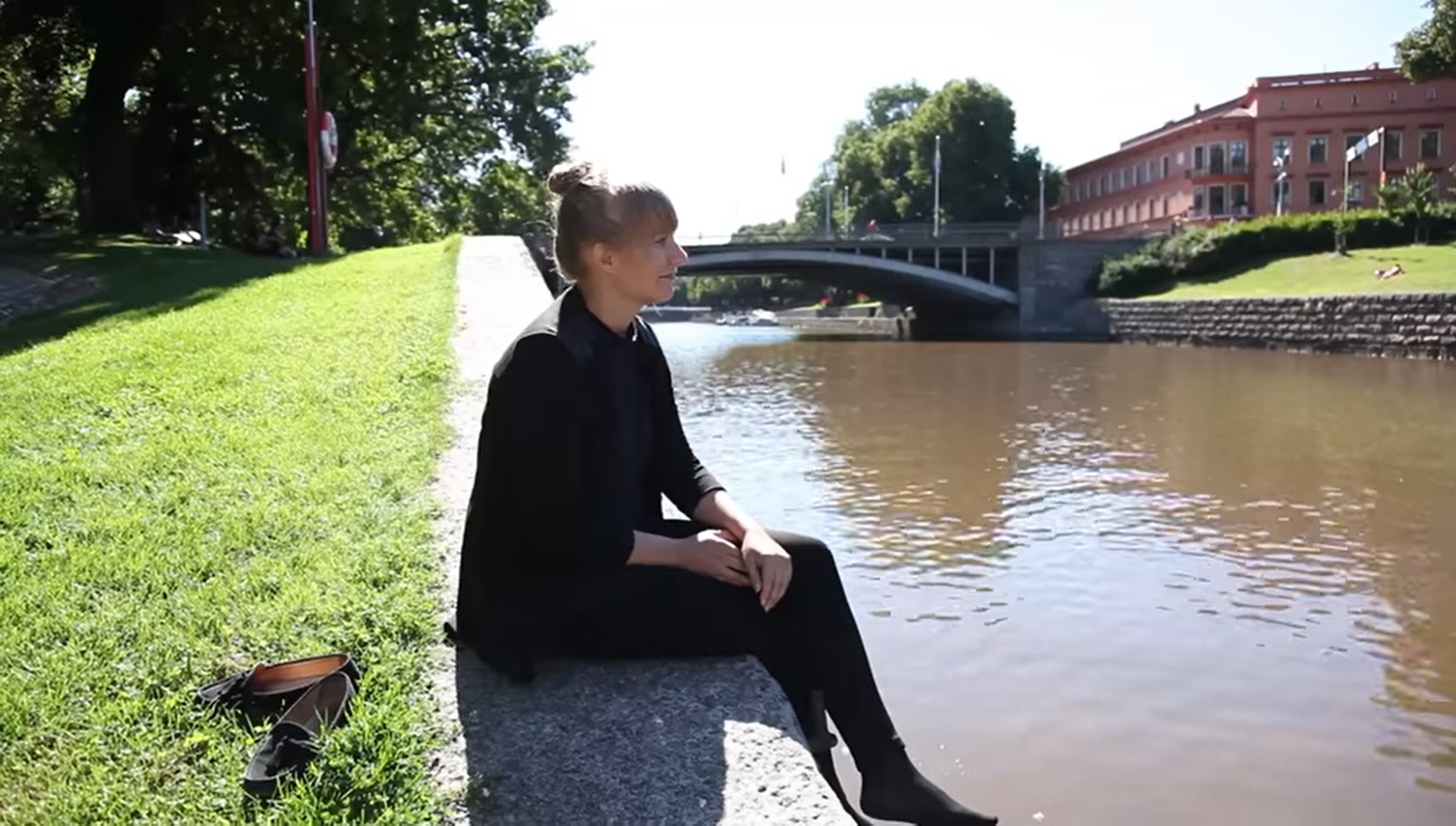}}\\
\subfloat[]{\includegraphics[width = 1.5in]{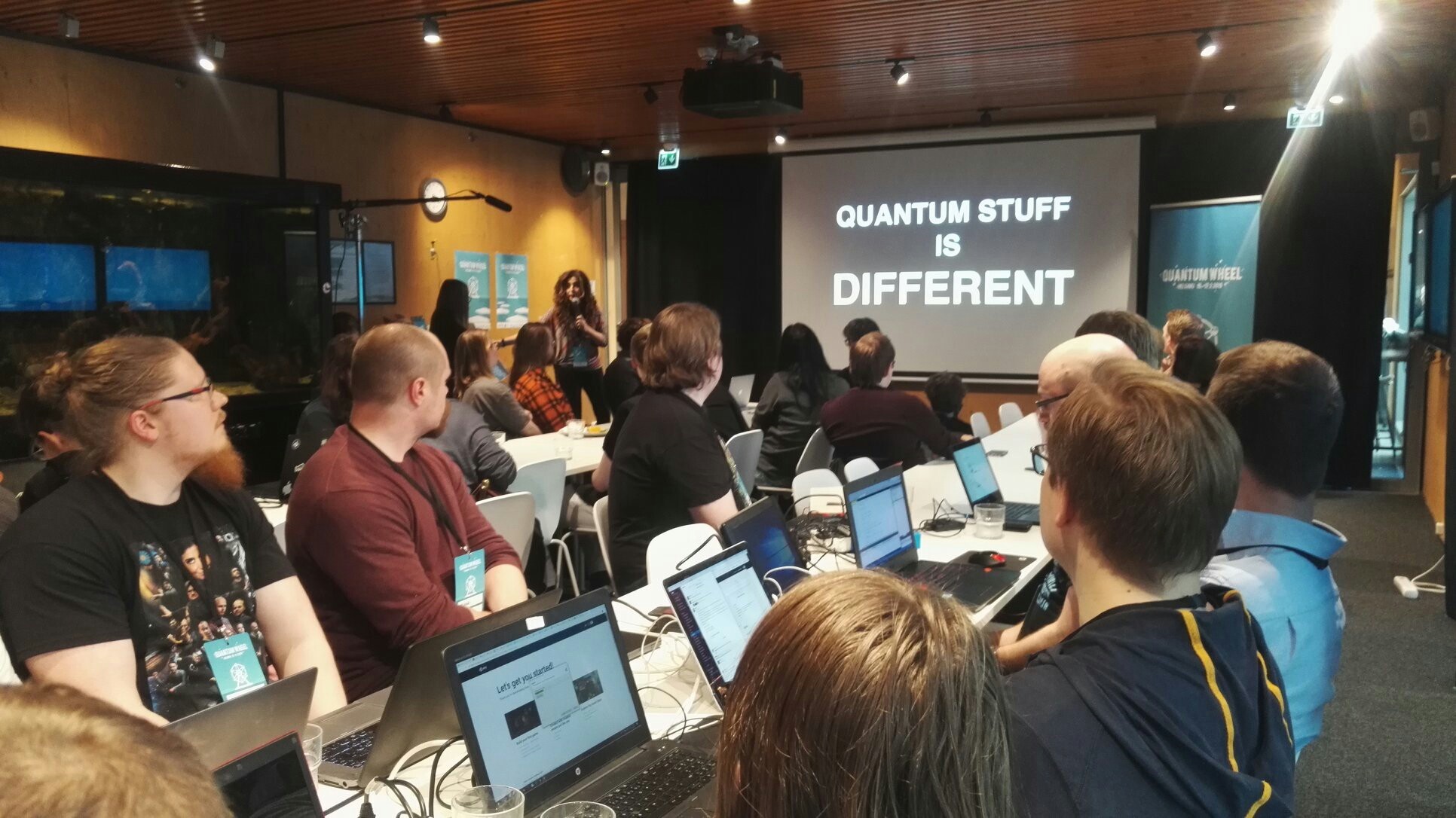}}
\subfloat[]{\includegraphics[width = 1.5in]{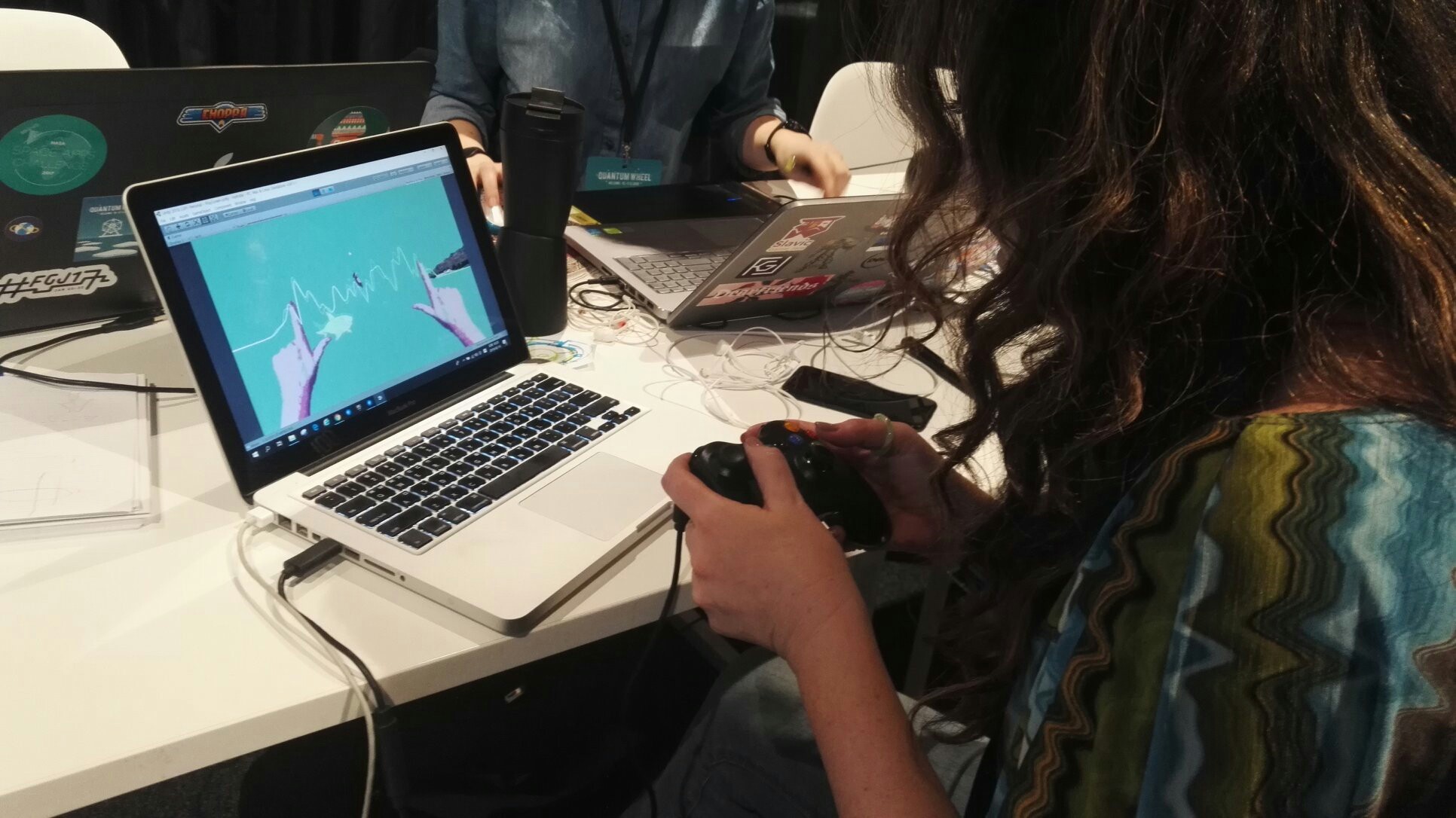}} 
\caption{Pictures from Quantum Game Jams. a) A picture from the Tuorla observatorium dome at the first Quantum Game Jam. \textit{Jaakko Vainio}, 2014 b) A screen shot from a theme presentation video from 2015, \textit{Turku Quantum Technology} c) A picture from the 5th Quantum Game Jam, Quantum Wheel in Helsinki; the presentation on a design constraint by Professor Maniscalco. \textit{Laura Piispanen}, 2019. d) A picture from the 5th Quantum Game Jam, Quantum Wheel in Helsinki. Professor Maniscalco testing the game Hamsterwave. \textit{Laura Piispanen}, 2019.}
\label{fig:places}\vspace*{-6pt}
\end{figure}

QGJs were modeled based on the experiences of Global Game Jam (GGJ) in Finland \cite{kultima16b}. The events took place over the weekends resulting in about 48h of development, including rest. The participants created games in teams and based on design constraints that were presented in the beginning of the events. The events were designed as non-competetive, in order to facilitate playful attitude and creativity \cite{goddard14}.
QGJ was organized five times: in 2014, 2015, 2016, 2017 and 2019. The setting of the QGJ changed from year to year including different locations, technologies and design constraints, as well as how participants were recruited to the events. In 2014 the event was run only in Finland (See Figure \ref{fig:places}a), in 2015 there were five physical sites (Finland, Denmark, USA and Brazil) and the possibility to participate online. In 2016, there were five physical locations, four in Finland and one in Georgia. In 2017 the jam was opened for anyone to participate online through Itch.io in addition to the physical location in Finland.
As a final event on the QGJ project, an upgraded event, Quantum Wheel took place as a single-location game jam in Helsinki, Finland in February 2019 (See Figures \ref{fig:places}c and \ref{fig:places}d). In addition to the physicists and game experts from the organizing parties, experienced jammers and game developers were invited to the event and further participants were recruited via applications. Quantum Wheel was the first in the series to utilize actual quantum computers. 

\begin{table*}
  \centering
 \includegraphics[width=0.8\linewidth]{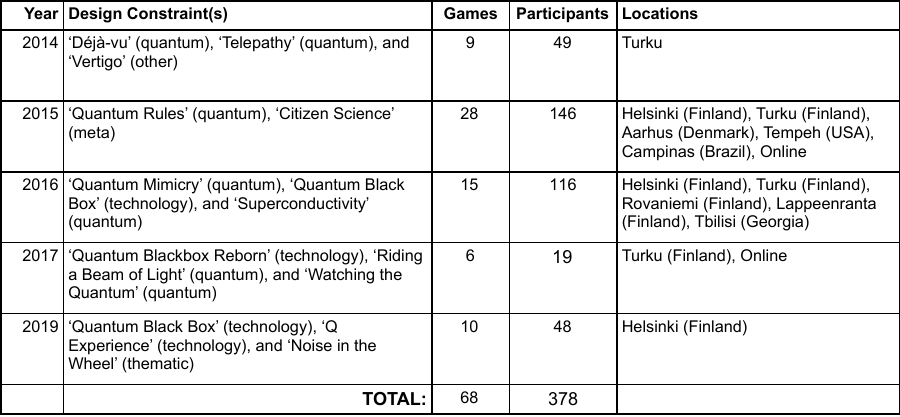}
  \caption{Design constraints, games, participants, and locations of Quantum Game Jam 2014-2019}
\label{tab:design}
\end{table*}

\subsubsection{Design Constraints}
In many game jams, the participants are given a theme as a design constraint. For example in GGJ 2021 the shared theme was “Lost \& Found” \cite{ggj21}. Kultima et. al \cite{kultima16} have been exploring how jam themes can be interpreted as design constraints and how they can be framed depending on the approach of the design process \cite{kultima15a}. In general terms, design constraints do not always mean limitations, but necessary starting points for design, as without them the design has no clear purpose or direction \cite{lawson06}. Design constraints can come from legislation, materials, clients or for instance from the personal preferences of the designers \cite{lawson06}. 

One of the characteristics of QGJs was that the participants were able to choose between the external design constraints \cite{kultima16}. The events had two to three constraints catering to different interests and skill levels (See Table \ref{tab:design}). Some of the constraints were thematic allowing open interpretation, some technical providing assisting tools and technologies, and one constraint was inviting participants to create citizen science games (2015). At the main locations in Finland the design constraints have been presented in person by the organizing physicists (See Figure \ref{fig:places}c), but they have been also communicated through short videos to provide access to other participants (See Figure \ref{fig:places}b). For the first event, the jammers were also offered a theme that was not directly about quantum, but another topic in theoretical physics accompanied with open data of galaxies.

While the QGJ design constraints were designed to inspire and guide the participants towards the goals of the events, there was also room for the self-imposed constraints \cite{kultima16} of the participants catering to the intrinsic interests towards quantum mechanics and quantum technologies, or other provided affordances of the events.
\enlargethispage{13pt}
\subsubsection{The Locations}
As the original aim has been to attract experienced game makers and jammers \cite{kultima19}, it has been important to create exceptional jam setups. In the first year of QGJ, the jam was held at an astronomical observation station. One of the attractions of this jam was the possibility to create games utilizing the full-dome projector of the observatory’s planetarium (See Figure \ref{fig:places}a), but also the otherwise-limited access of the venue. The second jam continued with a similar attraction hosting the jam at the Heureka science center in Vantaa, Finland, where also access to a full-dome projector was provided. In 2016, the jam took place on the Otaniemi campus of Aalto University, where jammers were advertised to have access to the quantum technology laboratories of the campus.

The 2017 location was more conventional: the participants were hosted at a cowork space, where other game jams had been hosted locally. However, in the case of the fifth event, The Quantum Wheel, the location was again exceptional: a Ferris wheel and an outdoor pool (in freezing cold weather) with a sauna was available for the jammers. Inspiration for the theme “Noise in the Wheel”  was sought from riding the Ferris wheel on the second day of the event. The attractive environment was chosen to draw participants to apply, but also to create the best possible atmosphere for multidisciplinary work and seeking of exceptional ideas. 

\subsubsection{The Participants}
The recruitment practices to the events varied from year to year. The first four events have had an open participation for anyone interested and the possibility to run “satellite sites” was offered in 2015, 2016 and 2017. 

The fifth jam was built as an invitation-only event to ensure experienced participants. The recruitment was done through applications and invitations, bringing in such experienced game developers as the designer of \textit{Angry Birds}, Jaakko Iisalo and creator of \textit{Baba is You}, Arvi Teikari together with other talented game experts from around the world.

The participant numbers of the QGJ events have been between 19 and 146 (See Table \ref{tab:design}). Our detailed participant information archives were slightly incomplete as not all the jams collected the same amount of information on the participants. However, based on the retrieved registration information from 2014, 2015, 2016, and 2019 events and on the credits of the games produced in 2017, we could infer that at least 22 participants took part in the events more than once. Out of the 313 registrations, we had the gender information of 252 participants. Of them 25\% were females, 75\% males. Special attention on diversity was given on the curation of the participant list of 2019 QGJ, concluding to 33\% females and 65\% males, 1\% other. Also the home countries of the participants were recorded from the fifth event: the majority of the participants were from Finland (60\%), but the event had also participants from Germany, Italy, Russia, Estonia, Austria, USA, and South Korea.\enlargethispage{13pt}

\subsubsection{The Organizers}
The game expertise and access to the game jam communities has been provided by the Finnish Game Jam organization, known for their experiments in game jams and active engagement in the GGJ community \cite{kultima16b}. The research team of Turku Quantum Technology (TQT) has offered their expertise in quantum mechanics by participating in the jamming events, and working on theme design. Some of the participants of the events in teams were also from the organizing teams. There were about 1-4 organizers in each event that were not participating in the game projects. 

\subsubsection{Games}
In game jams, the participants aim to create something playable, instead of mere concept descriptions or ideas. However, typically a game jam results in small games that can be more described as prototypes than full-fledged games \cite{decker15}. 

Altogether 68 prototypes were created as part of the QGJ events between 2014 and 2019 (See Table \ref{tab:games}). In 2015 28 games were created, while the other events resulted in six to 15 games. Most games from the jams have been uploaded online through Itch.io. Only the 9 games from the first event in 2014 are not available online anymore, but were retrieved in order to include them in this study. 

\subsubsection{Additional Tools}
Through the years of QGJ, also some tools have been provided to ease the creation of science games. A specific tool, Quantum Black Box (QBB),  was created by the TQT research team to help facilitate the creation of citizen science games. The tool was introduced at the 2017  jam and further developed based on the experiences from the jams. QBB simulates a certain quantum optimal control problem actively researched by the research team. QBB was intended to be used in game projects that aim to become citizen science projects and designed to allow the  development time to be used on the game design instead of quantum physics. 

QBB was originally written in Python and provided as such for the jammers. For the 2019 event, QBB was provided as a Unity plugin with an example game with source files. Eleven out of the 68 games from the QGJ events used QBB and it has since been further developed together with a commercial game company MiTale powering a citizen science game prototype, \textit{QWiz} \cite{qwiz}.
Furthermore, in the 2019 jam, there was also a set of creative tools provided by IBM aimed for the teams working on quantum computers \cite{qpiano,collapse, qiskit}.

\subsection{Online survey}
For the purpose of this study, we designed an online survey for the participants of the fifth jam. After observing four jams, we were interested in understanding in detail the potential challenges that a science game jam might have. Specifically we were interested in what kind of challenges the jammers faced and how iterative the processes were.

The survey consisted of seven questions, out of which four were open questions. The questionnaire was sent to the participants of the jam on the last day of the event in emails and posted also on the Discord server of the jam. The questionnaire was anonymous, but some background questions were inquired (See Table \ref{tab:roles}): which role the participants took and if they were contributing to one or more teams. We got 15 responses out of the 48 participants of the Quantum Wheel 2019 event.

\begin{table}
  \centering
  \includegraphics[width=\linewidth]{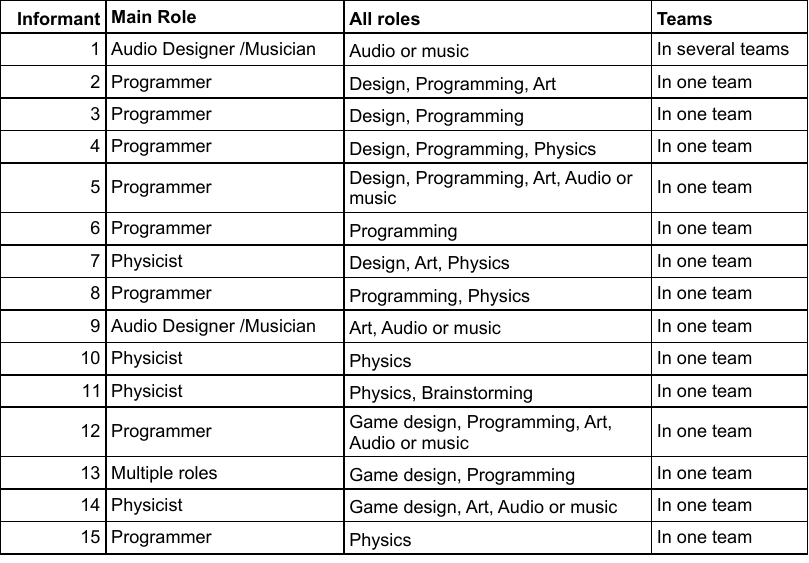}
  \caption{Team roles of the survey respondents.}
\Description{Team roles of the survey respondents.}
\label{tab:roles}\vspace*{-6pt}
\end{table} 
Unfortunately, we were not able to collect survey data from the first four events, but there was a small overlap in the participants: six out of 48 participants of the Quantum Wheel had taken part in several QGJs, and three participants had been taking part in all events.

\subsection{Game Analysis}
\begin{figure}
  \centering
 \includegraphics[width=1\linewidth]{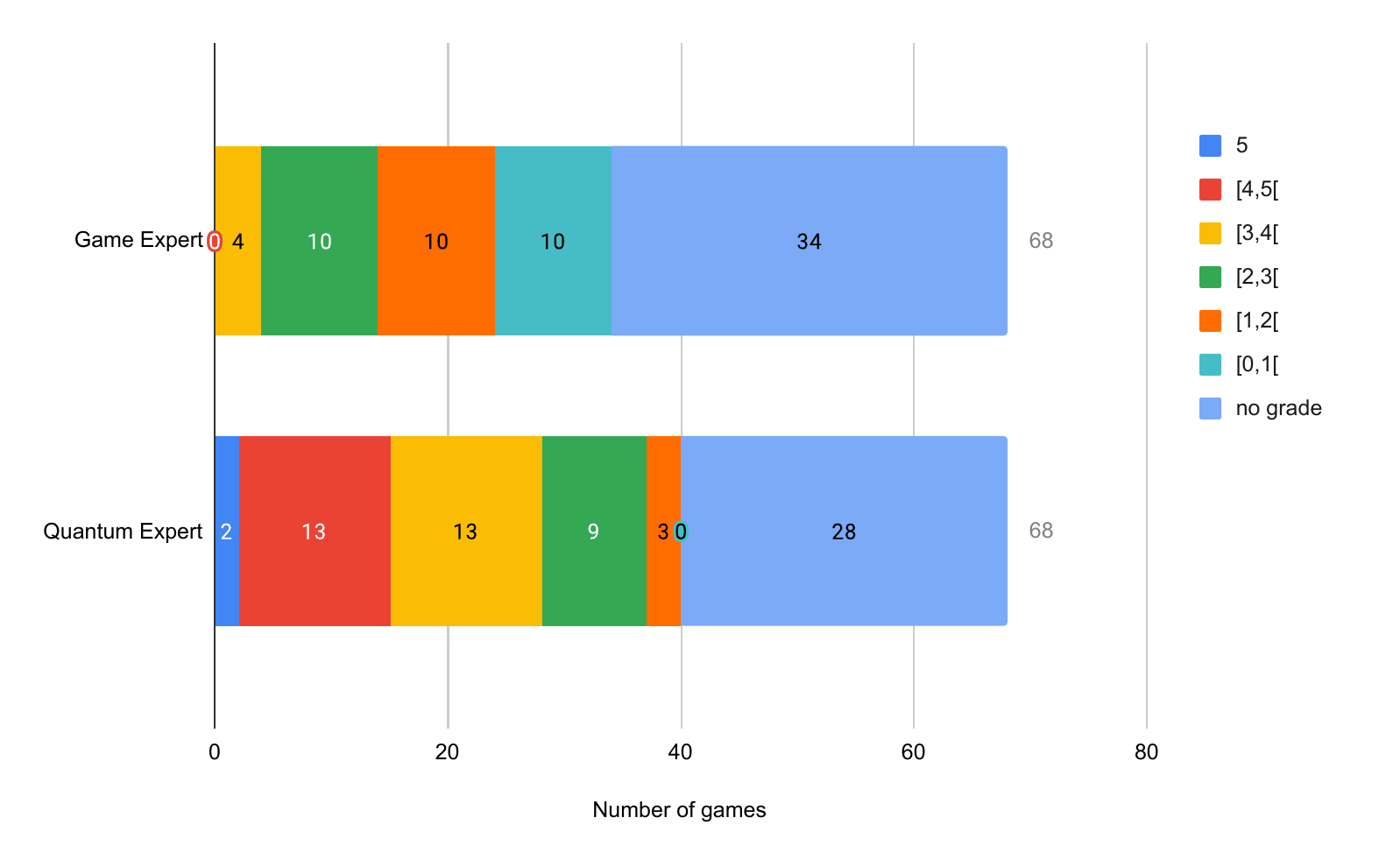}
  \caption{Grades of the gameplay.}
\label{tab:gradesG}
\end{figure}

In order to evaluate how well QGJ worked as an open innovation method for directly producing feasible prototypes for science games, we performed an analysis on the games created in these events since 2014. 
Out of the 68 prototypes created in the jams, 66 had a working prototype or other materials that we were able to use in the analysis. We set a form to facilitate the evaluations and two authors of this paper were assigned to test the jam games for a set time period: 15 min. maximum of playing the game, and 15 min. maximum for installation. The experts then evaluated the given games in terms of usability, enjoyment and creativity. Games were further evaluated for their potential to be developed into citizen science or (quantum) educational games. The evaluations were made on a scale from 0 to 5, where 0 and 5 corresponded to given statements (See Tables \ref{tab:statements}, \ref{tab:gradesG}, and \ref{tab:gradesQ}). As the jam games were not fully developed products, we devised a statement batch that was focused on the perceived potential of the games.

Out of all the games produced at the jams, at least 45 games were playable on PC. There was also one card game and one board game, five VR games, three games for full-dome projectors, one multiplayer game requiring server access and one game was designed for mobile phones with an accelerometer. These games were not evaluated on their gameplay due to the extended need for setup.

One evaluator is a PhD candidate in physics, working on the research of quantum technologies, with minors in game design and pedagogy and another major in computer science (Quantum Expert). The Quantum Expert was involved on the fifth Quantum Game Jam as a participant, and has also been involved in other citizen science game prototyping processes. The other evaluator is a lecturer in game design with Master’s degrees in Arts in New Media and Computer Science (Game Expert). The game expert has industry experience in game design but no education in quantum mechanics. The Game Expert also has game jam experience from multiple jams but has not participated in Quantum Game Jams like the Quantum Expert.

\begin{table}
  \centering
  \includegraphics[width=\linewidth]{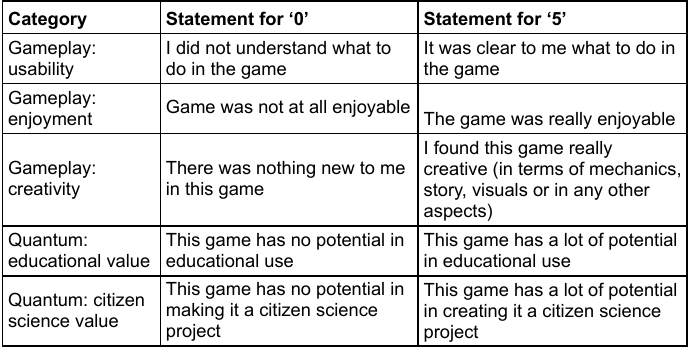}
  \caption{The statements corresponding to the values 0 and 5 on the scale used on the expert evaluations.}
\Description{}
\label{tab:statements}
\end{table}

The total number of the games evaluated was 61. The experts evaluated the games independently and the work resulted in 92 evaluations. From these, only 73 were based on playing the game as 45 games were available on PC at the time of the evaluation without requiring extended setup time. In addition, 16 games were evaluated only on their quantum potential through videos and other available material, deemed enough to be included in the analysis. Some prototypes were familiar to the Quantum Expert, as she took part in one of the jams.
The educational and citizen science potential was evaluated only by the Quantum Expert, whereas the gameplay evaluation was conducted by both. From the grades of the educational and the citizen science potential a simple average was calculated to constitute a single grade, referred here to as the “Quantum grade”. Similarly the grades for gameplay were averaged into one number.

\begin{table*}
  \centering
 \includegraphics[width=0.85\linewidth]{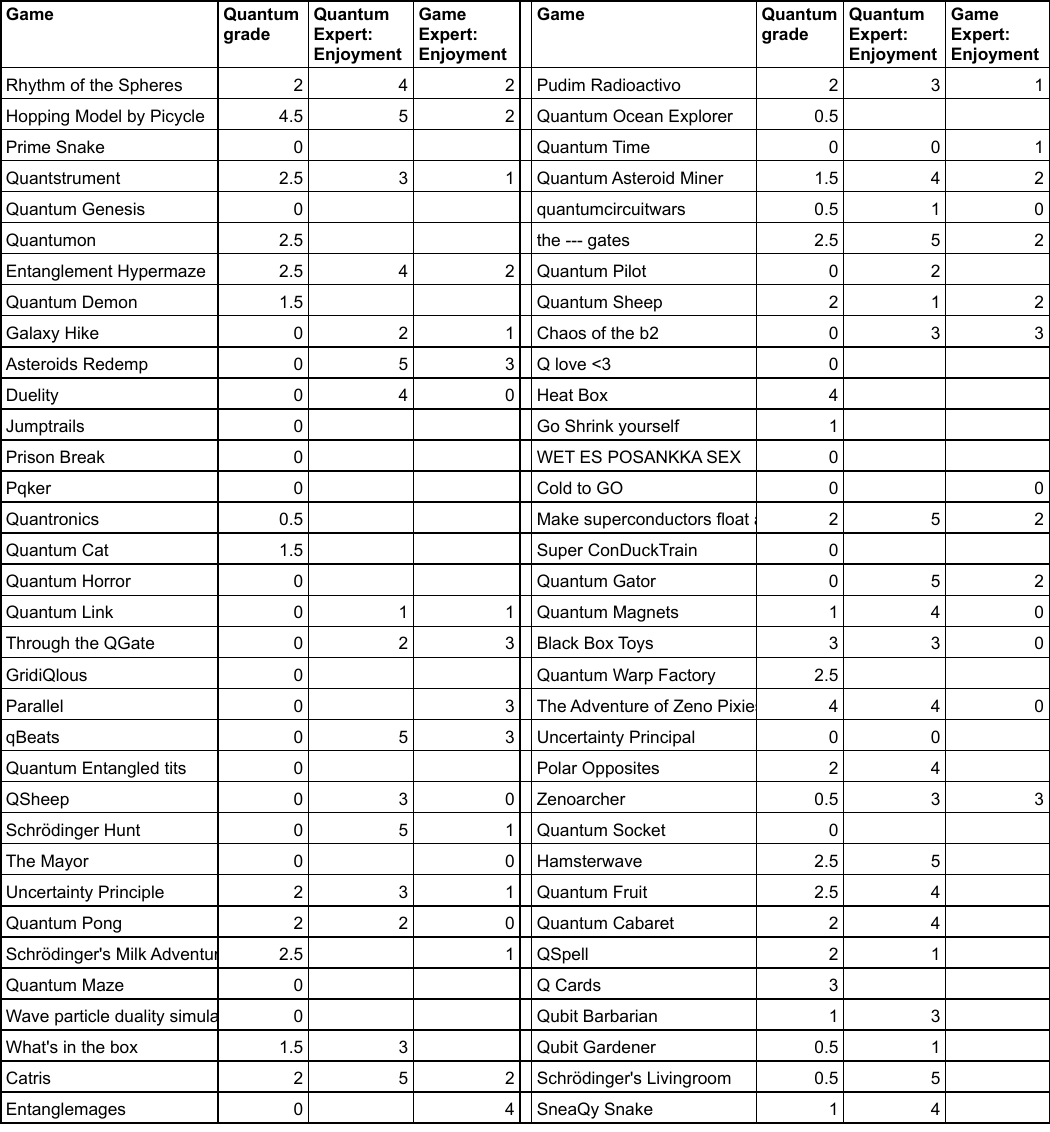}
  \caption{List of grades evaluated for all the games available from Quantum Game Jams between 2014-2019.}
\label{tab:games}
\end{table*}

\section{Results}
\subsection{How QGJs worked as a method for directly producing science game prototypes?}
\begin{figure}
  \centering
 \includegraphics[width=\linewidth]{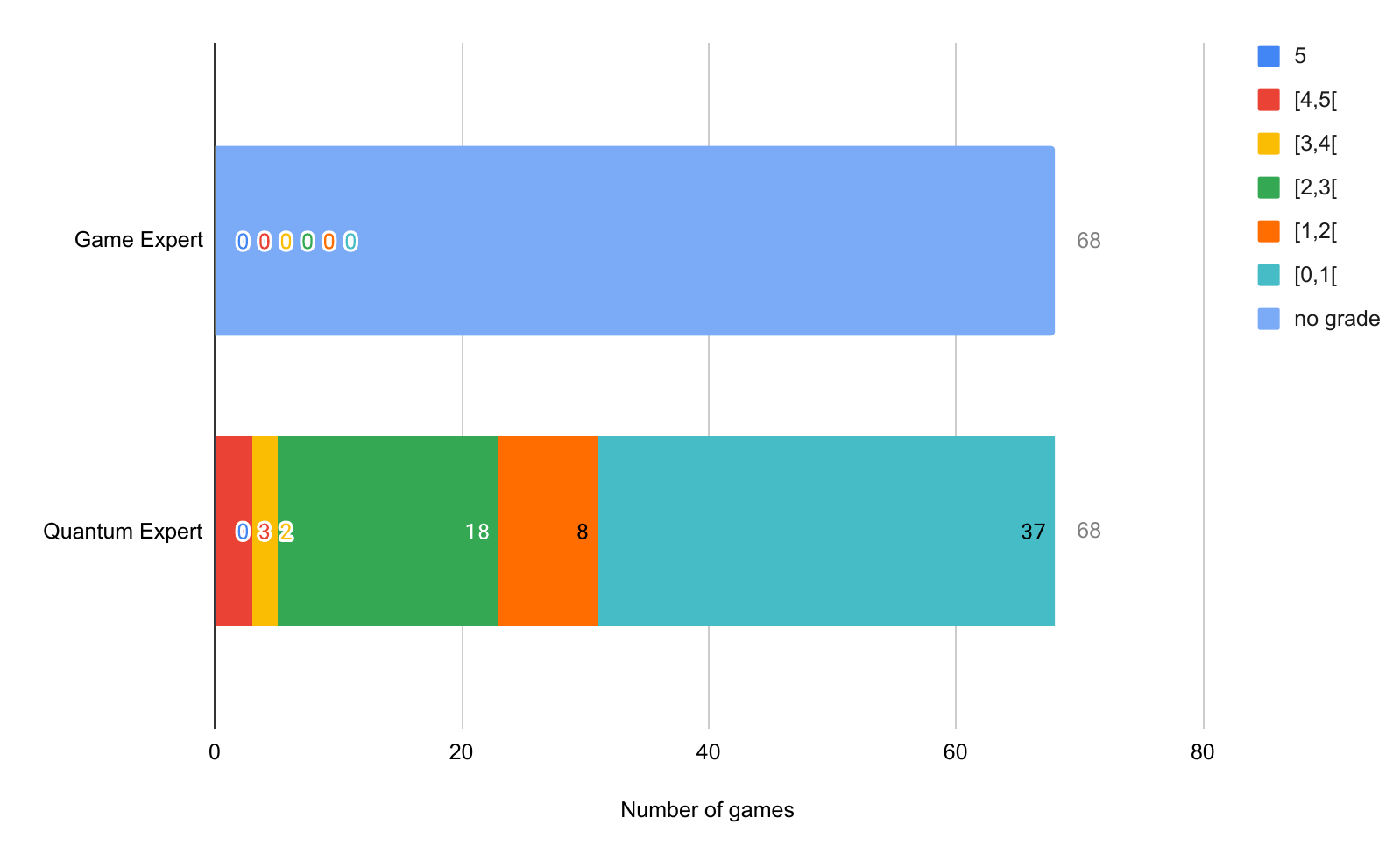}
  \caption{Grades of the quantum aspect.}
\label{tab:gradesQ}
\end{figure}

Out of the evaluated 61 games, only few games were marked by the Quantum Expert as having potential in quantum educational or citizen science projects (See Tables \ref{tab:games} and \ref{tab:gradesQ}). However, the Quantum Expert was evaluating the games higher in usability, enjoyment and creativity than the Game Expert. Game Expert’s average enjoyment grade for the games was 1.5, whereas for the Quantum Expert it was 3.3, and only 7 of the games got the average grade 3 or more on the gameplay from both experts.
The quantum themes were incorporated into the prototyped games on different levels. A couple of games rely on a direct representation of quantum machinery or a laboratory setup. One such example is \textit{The --- Gates} from 2016 that is a set of puzzles based on the principles of quantum optics and the tools used in a laboratory. Another such example is the card game \textit{Q|Cards>} from, 2019, where players set their cards on the table in order to form a so-called quantum circuit, that is widely used to represent the operations in quantum computing.

\begin{figure*}
\subfloat[]{\includegraphics[height = 1.84in]{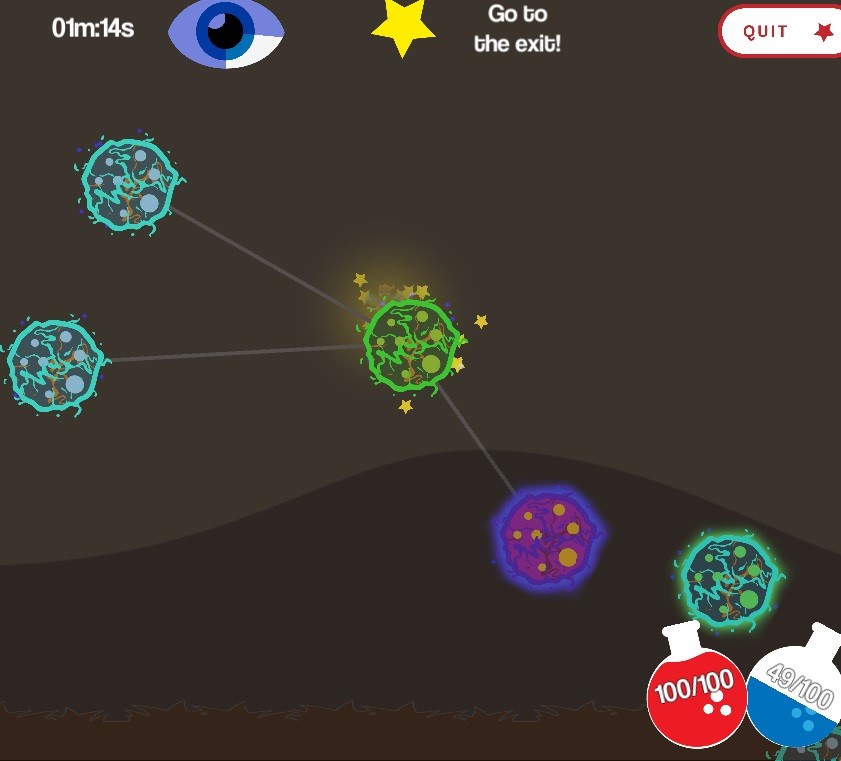}} \hspace*{10pt}
\subfloat[]{\includegraphics[width = 2.5in]{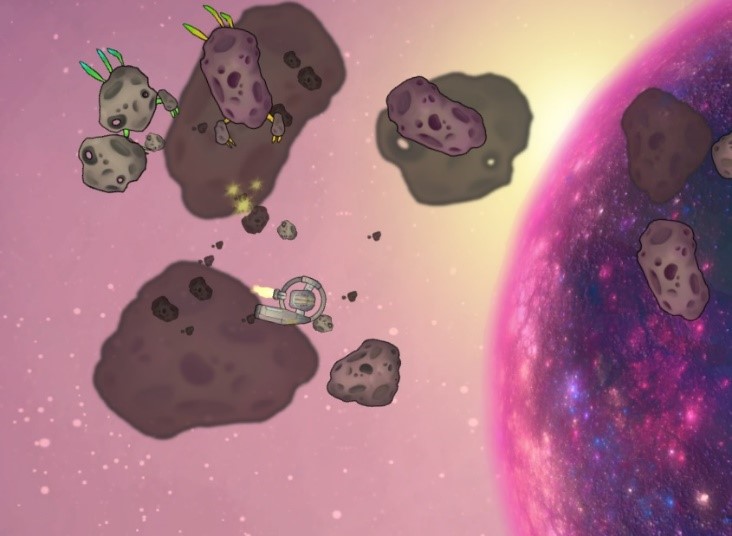}}\\
\subfloat[]{\includegraphics[width = 2.51in]{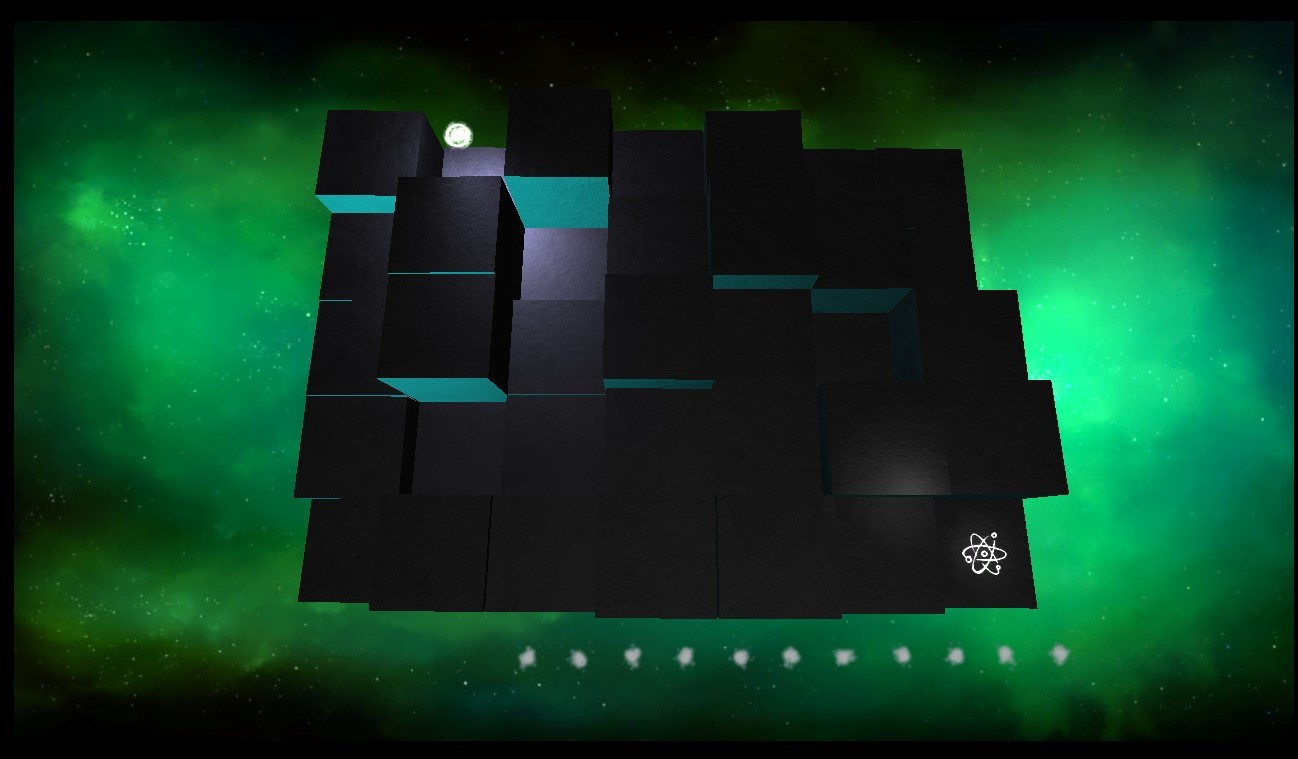}}\hspace*{10pt}
\subfloat[]{\includegraphics[width = 2.51in]{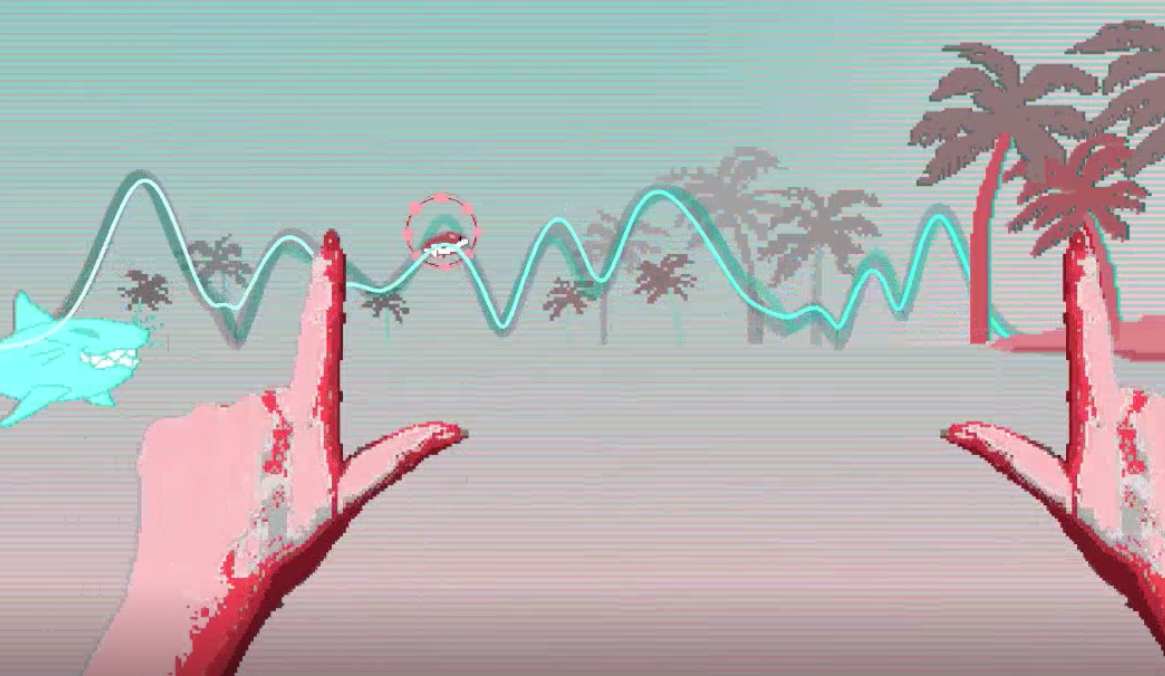}} 
\caption{Screenshots from games produced in Quantum Game Jams: a) \textit{The Adventures of Zeno Pixies}, 2017 b) \textit{Quantum Asteroid Miner}, 2016 c) \textit{The Hopping Model by PiCycle}, 2014 d) \textit{Hamsterwave}, 2019 }
\label{screenshots}
\end{figure*}

Some games have the connection to quantum physics slightly disguised. These games might for example have a clear visual\break presentation originating from the researchers, adding graphics, story and animations on top of it. As an example the game \textit{The Adventures of Zeno Pixies} developed in 2017 (See Figure \ref{screenshots}a) is a puzzle game, where the player learns about the quantum Zeno effect and quantum entanglement by guiding little pixies through a colorful network. The underlying mechanics of the game rely directly on a model of a so-called quantum random walker on a quantum network, concepts that refer to a theory used in research areas such as quantum computing and quantum communication \cite{kempe03}. The team had two physicists from TQT, who were experts in this area of research. Although it has so far been left on a prototype level, it manifested potential to be developed further to incorporate complex quantum networks, which are a topical issue in quantum information sciences \cite{kendon06,kozlowski19,dahlberg19}.

A lot of games incorporated the ideas from quantum phenomena to a set of clear rules for the game, but the effect itself might not be based on actual calculations. A character might be able to switch between different forms, that represents the act of being in a quantum superposition. Or an obstacle might reveal its true shape only when it is close enough. Such functionality is usually added to represent the act of a quantum measurement. One such example is \textit{Quantum Asteroid Miner} from 2016 (See Figure \ref{screenshots}b), where the aim is to mine through asteroids that might end up turning into enemies. This creates a thrilling atmosphere and relies on the act of surprise through randomness that is inspired by the statistical nature of quantum physics phenomena. Many of the games do exactly this, use the given tools just as random number generators or include a classical random number generator to the game to express the seemingly chaotic nature of quantum phenomena.  

In a more hectic game the quantum side might at first seem a bit more hidden, as the player has to focus on the rapidly changing situations. This is the case for example in the \textit{Uncertainty Principal}, 2017, where the player acts as the principal of a daycare and needs to be sure to check different parts of a room, and always finds new accidents about to happen. Many games made during the jams are utilizing the probabilistic nature of quantum mechanics: a certain feel of surprise and chaos is present in many of them, even though quantum mechanics itself is not chaotic.

Not all the games had documentation about the aspects of how quantum physics was incorporated in them. One such game is \textit{Hopping Mode} by PiCycle from 2014 (See Figure \ref{screenshots}c). The additional credits in the project include quantum physicists, so a connection to quantum physics is probable. Nevertheless it was rated high on the quantum grade as the Quantum Expert saw an informative representation of quantum phenomena in it and felt there would be a possibility to combine the game with a specific numerical simulation of a quantum mechanical system. 

It seems that very few jam games excelled both in gameplay and in quantum evaluations. This is also supported by our observations throughout the events. Oftentimes the games that were more entertaining lacked in accuracy and the ones that had potential to depict physics phenomena lacked in gameplay. Combining the subject matter with fun game mechanics without one or the other suffering is considered one of the main challenges in serious games design in general \cite{marfisi10}.

Even if the games would have potential for full-fledged science games, the further development of the games can be challenging for practical reasons. The teams might not have time to continue the work after the jams and the prototypes would require hours of work to make them full games. Despite this, from the final event, at least three games have been developed further. 

\textit{Hamsterwave} (See Figure \ref{screenshots}d) was further developed for a game exhibition at Finnish library Oodi as part of a showcase of Aalto University, the card game Q|Cards> was further developed and released in collaboration with IBM and MiTale, and \textit{QSpell} has been the inspiration for a virtual reality citizen science game prototype QWiz \cite{qwiz}. Furthermore, as a result of all the jams, the QBB has been developed further to offer a refined platform for commissioned works, such as \textit{Quantum Garden} \cite{quantumgarden}. These can be considered as an indirect impact of QGJ.

\subsection{How QGJs worked as a format for interdisciplinary collaboration?}
One of the main challenges has been facilitating the process of collaboration between the game discipline(s) and the physics discipline. There is certain friction between the design goals of both parties: while quantum physicists often aim to depict scientific phenomena as accurately as possible, the game makers aim to translate these into enjoyable and fantastic experiences – in a very short time span of game jams. In some game jams the physicists were drawn into groups that had more physicists, but less game designers or they were working as designers themselves – and in some the physicists were not fully involved in the design process for the entire time of the jam. The teams that were more balanced, had previous QGJ experience, or had overlapping expertise seemed to be more successful in making both enjoyable and feasible prototypes. 

\subsubsection{Iterative development}
A key component in making good games is the iterative nature of the design process \cite{kultima15b}. It is not very clear how much projects evolve within the framework of game jams. Further understanding of how much iteration is involved, could help us build better impact for science jams.

From the survey data, we were able to gather some examples of iteration in the QGJ design processes. Some of the games had changed their genre or core of the game experience (for instance from a puzzle platformer to a “surfing game”, from a platformer to a narrative driven game, or from a competitive game to a collaborative game) throughout the process, but some teams merely added a few features. Some did not manage to implement all planned features or did not feel that they met the quality of the game that they initially planned for. Some of the shorter responses just mentioned that there were no major changes and a couple of respondents expressed that they did not understand the idea to begin with, thus were not able to iterate on the design idea. 

The format of the QGJ is quite short for an iterative process and further improving on the game concepts through prototyping would improve the impact. If a game has less room for changing the core design and utilizing design opportunism, the more iterations might be needed \cite{kultima18a}. Science game jams seem to especially need space for further iterations and the format could take this into consideration.

\subsubsection{Challenges} 
Most of the challenges reported in the survey had to do with technology. Problems with setting up the \textit{Qiskit} framework, used to access to IBM quantum computers, or not understanding what the QBB was meant to do were mentioned. Some had technology issues that were not related to quantum: one person had problems with their laptop and one respondent expressed that their team did not have enough experienced game developers to meet their initial vision of the game.

\looseness-1 Communication issues were also reported: there were a couple of non-physicists reporting that the physicist in the team lacked in understanding how games are made or how brainstorming works. It was reported that the physicists in the team had disrupted the brainstorming by criticizing ideas too early and another respondent suggested that the physicists could have also had some creativity training in the beginning of the event to join to the teams with more self-confidence, basic skills and being able to work independently (on the game development project). On the other hand, some physicists also reported that it took long to explain some issues in physics that were used in the games, and some respondents expressed that they did not either have enough time to learn the physics or ended up not understanding what the game was eventually doing physics-wise.

Through observation, we also witnessed some of these struggles. One team particularly had social challenges in the beginning. It took long for them to settle the group and agree on the vision, but eventually they managed to solve their issues. Some physicists were not able to attract enough experienced game developers to their teams (or did not find other ideas appealing enough to join the other teams) and as a result, their games lacked for instance visual appeal. Some teams did not have physicists involved in their teams at all, resulting in games that were not connected to physics on a deeper level, i.e. using the quantum computer for more than just as a random number generator or displaying the quantum phenomena in a way corresponding to theory. 

Reflecting on the results from the survey and analysing the games, we propose ways to improve the interdisciplinary communication: adding additional day(s) for both parties (developers and scientists) to work on the basic understanding of game development and game jamming as a process as well as basic understanding of quantum phenomena. A workshop in game making in the beginning of the jam should also include basic topics, such as how creative teams work together in game making communities: brainstorming techniques and creative decision making, selling one’s ideas, motivating the team members etc. Technological issues should be solved in the development of the kits that are provided for the participants and the team formation could be overlooked better to mix the disciplines and match the expertise more evenly. Instructing the teams to pay attention in explaining their prototype (in their documentation), could also improve the path of the project to further developed games.

What we have been able to witness throughout the years is that QGJs has helped the scientists formulate their research questions to fit game making and network with game developers. Altogether, what we also found out is that usually the more experienced (quantum) jammers were able to tackle the design constraints in an interesting way providing prototypes that had greater potential. The series of jams has made it possible to form a particular design expertise.
\vspace*{-6pt}

\section{Discussion}
The goal for organizing QGJs has been to explore and produce prototypes for science games, especially quantum games. In science games, both the enjoyment of the gameplay and scientific usefulness of the games are important. An ideal science game needs to be attractive and engaging as well as able to assist in the scientific goals.

Many serious games with scientific agendas aim to attract large masses and retain players to work for the science projects, as is in the form of citizen science games. If such factors are not covered in the early phases of the design, they might be hard to introduce later in the production process \cite{kultima18a,fullerton14}. In particular, for educational games and citizen science games research has been made for developing rigorous guidelines in designing such games \cite{mayer14, defreitas06, marklund}. This lesson seemed to be one of the most difficult ones to convey to the physicists of the projects – the appeal factor is not something that can be glued on later when the physics are in place - it is intertwined within the design process. Furthermore, mechanics-based problems in game projects require more iterations, which is of course also central to all game development \cite{kultima15b}.  While issues with usability can be solved by further developing the game prototypes, it also takes a lot of time. Even though the QGJ project seems to have some success in finding prototypes that meet both potentials, each project needed closer examination and further work.

Furthermore, the QGJs have been places for learning game development \cite{guevara, macklin, preston}. Physicists have improved in communicating with game developers, not only by learning game development vocabulary and gaining personal experience in the field, but also becoming better at explaining their own expertise. However, it is notable that the experience of the physicists can then be limited to game jams only, which in return provide a biased view on the realities of game production. In further development of some of the projects, it has become a challenge to explain how many additional hours next steps in development can take, what kind of changes can be done and how big actual production costs might be. Game jam prototypes sometimes seem ready for distribution. Nevertheless, their code, additional content and functionalities such as menus, narratives and characters take much longer in development. Above all there are usability issues, debugging and all other important factors of creating working standalone products that retain users, that should be addressed. Game jams can be perfect platforms for non-makers to learn more about game development but can also create something we call the “jam bias”. If development experiences are limited to jams only and reinforced by a series of them, the jam bias can be hard to break. Paying specific attention to this already within the first jam setting can alleviate potential future challenges.

As an improvement for the process of science jams in general, we suggest that the method is tweaked beyond the archetype of a GGJ. There seems to be a need for educating and framing the disciplinary groups beforehand of the jam to improve the results of the projects. Such a model has been partially utilized in QGJ in the form of science presentations, but the length of these presentations has been inadequate. There is a need for a full day workshop similar to the one in Sami Game Jam \cite{kultima19, laiti}. However, the workshop needs to cover also principles and practices of game making to prepare the domain experts and non-creatives for better collaborative work --- as long as the events also aim for full participation of the content experts.

\section{Limitations and future work}
It is challenging to evaluate jam games outside the events. As the games are not well documented and there are a lot of usability issues, the jam games might not always make sense for the player without the creators explaining the project. The analysis concentrated on externally perceived potential of the prototypes, which could change if the projects were further polished or explained face-to-face. We also had limited resources for the expert evaluators. In future work, this could be improved by recruiting more experts to see how the evaluations would vary. 

We also think that it would be important to organize more science game jams, to test the hypotheses formed from the reflections of the results. Even though we have hopes that our lessons could improve the method, it is not given that these improvements would dramatically improve the rate of useful prototypes. The indirect impact of the method, then again, is difficult to measure.

\section{Conclusion}
In this paper, we have evaluated the method of particular science game jams between 2014 and 2019. The series of Quantum Game Jams have aimed to engage quantum physicists and game makers in interdisciplinary collaboration. The series of five events resulted in 68 game prototypes covering various aspects of science games and a specific software for enabling the development of citizen science games. At least three games have been taken further in development, the events have provided places for networking and both parties have learned from the expertise of the others. This project shows that game jams can provide platforms for interdisciplinary collaboration as well as places for creating science games. However, specific challenges in collaborative work need more attention: there needs to be time for introducing the presented disciplines to the others and to facilitate the learning experiences of both participating parties. It is not sufficient to only concentrate on introducing scientific problems to game makers, but it is also vital to teach the scientists on the basics of game development processes allowing them to participate, frame their goals and work on further development of the projects. 

\begin{acks}
Acknowledgements to all the participants of Quantum Game Jams through out the years. It has been a privilege to meet you and to get to work with you!\\ The research has been partly funded by the use of Academy of Finland PROFI funding under the Academy decision number 318937.
\end{acks}

\bibliographystyle{ACM-Reference-Format}
\bibliography{references}


\end{document}